\documentclass{emulateapj}
\usepackage{natbib}
\usepackage{apjfonts}
\usepackage{epsfig}

\newcommand{\tq}{$t_{Q}$}
\newcommand{\tQ}{t_{Q}}
\newcommand{\etal}{et al.}
\newcommand{\NH}{$N_{H}$}
\newcommand{\Mdot}{\dot{M}}
\newcommand{\Lbol}{L_{\rm bol}}
\newcommand{\dEdt}{\epsilon\Mdot c^{2}}
\newcommand{\LB}{L_{B}}
\newcommand{\LBo}{L_{B,obs}}
\newcommand{\LBm}{L_{B,min}}
\newcommand{\Lcut}[1]{10^{#1}\,L_{\sun}}

\shorttitle{A Model for Quasar Lifetimes}
\shortauthors{Hopkins \etal}
\slugcomment{Submitted to ApJ Letters 01/28/05}
\begin{document}

\title{A Physical Model for the Origin of Quasar Lifetimes}
\author{Philip F. Hopkins\altaffilmark{1}, 
Lars Hernquist\altaffilmark{1}, 
Paul Martini\altaffilmark{1}, 
Thomas J. Cox\altaffilmark{1}, 
Brant Robertson\altaffilmark{1}, 
Tiziana Di Matteo\altaffilmark{2}, 
\&\ Volker Springel\altaffilmark{2} 
}
\altaffiltext{1}{Harvard-Smithsonian Center for Astrophysics, 
60 Garden Street, Cambridge, MA 02138, USA}
\altaffiltext{2}{Max-Planck-Institut f\"{u}r Astrophysik, 
Karl-Schwarzchild-Stra\ss e 1, 85740 Garching bei M\"{u}nchen, Germany}

\begin{abstract}
We propose a model of quasar lifetimes in which observational quasar
lifetimes and an intrinsic lifetime of rapid accretion are strongly
distinguished by the physics of obscuration by surrounding gas and
dust. Quasars are powered by gas funneled to galaxy centers, but for
a large part of the accretion lifetime they are heavily obscured by
the large gas densities powering accretion. During this obscured
phase, starbursts and black hole growth are fueled but the quasar is
buried. Eventually, feedback from the accretion energy disperses
surrounding gas and creates a window in which the black hole is
observable optically as a quasar, until the accretion rate drops below that 
required to maintain a quasar luminosity. We model this process and 
measure the unobscured and intrinsic quasar
lifetimes in a hydrodynamical simulation of a major galaxy merger. The
bolometric luminosity of the source is determined from the black hole
accretion rate, calculated from the local gas properties.  We
calculate the column density of hydrogen to the central galactic black
hole along multiple lines-of-sight in the simulation, and use these
column densities and the gas metallicity to determine the B-band
attenuation of the central source.  Defining the observable quasar
lifetime as the total time with an observed B-band luminosity greater
than some lower limit $\LBm$, we find lifetimes $\sim10-20$ Myr for
$\LBm=\Lcut{11}$ ($M_{\rm B}\approx-23$), 
in good agreement with observationally determined
quasar lifetimes. These numbers are significantly smaller than the
``intrinsic'' lifetime $\sim 100$ Myr obtained if attenuation is
neglected. We find similar lifetimes defined by an observed bolometric
luminosity greater than $10\%$ of the Eddington luminosity.  
The ratio of observed lifetimes to intrinsic
lifetime is also strong function of both the limiting luminosity and the
observed frequency.
\end{abstract}


\section{Introduction\label{sec:intro}}
The recent discovery of a strong correlation between the velocity dispersion of galaxy spheroids 
and the mass of their central, supermassive black holes \citep{FM00,Gebhardt00}
suggests that their evolution is strongly linked. One of the most fundamental parameters of 
black hole growth is the quasar lifetime \tq, which sets the length of rapid black hole growth.
%Quasar surveys in recent years have enabled detailed descriptions of many quasar properties, 
%and recently, a strong connection between galaxy formation and supermassive black holes
%has been discovered \citep[e.g.,][]{Tremaine02,FM00,Gebhardt00}, 
%suggesting that the evolution of both are linked.
%However, one of the most fundamental properties of quasars, the quasar lifetime, \tq, 
%remains highly uncertain. This is the critical timescale for supermassive black hole 
%growth and for the fueling and feedback mechanisms of active 
%galactic nuclei (AGN). The time that supermassive black holes radiate at 
%quasar luminosities determines the importance of a quasar phase for the 
%production of the present space density of dormant supermassive black holes and 
%is necessary in predicting and understanding both the quasar luminosity function and 
%the intergalactic infrared and X-ray backgrounds.
Observational estimates generally constrain the net quasar lifetime to the range
$\tQ\approx 10^{6}-10^{8}$ yr \citep[for a review see][]{Martini04}. 
These estimates are primarily based on demographic or integral arguments 
which combine observations of
the present-day population of supermassive black holes and accretion by the 
high-redshift quasar population \citep[e.g.,][]{Soltan82,HNR98,YT02,YL04}, or 
incorporate quasars into models of galaxy evolution \citep[e.g.,][]{KH00,WL02,DCSH03,Gran04}.
%or the reionization of HeII in the Universe (e.g. Sokasian et al. 2002). 
Recent results 
from clustering in quasar surveys \citep[e.g.,][]{PMN04,Grazian04} similarly suggest
lifetimes $\tQ\sim$ a few $10^{7}\,$yr. 

These estimates constrain only the total
time a supermassive black hole is radiating above a quasar luminosity threshold, and 
do not address a scenario in which a supermassive black hole experiences multiple
short, episodic quasar phases. A minimum episodic lifetime $\tQ\gtrsim10^{5}\,$yr is
required to explain the proximity effect in the Ly$\alpha$ forest \citep{BDO88}, while 
%and the sizes of ionization-bounded narrow-line regions \citep{Bennert02}, and 
multi-epoch observations give a minimum lifetime $\tQ\gtrsim10^{4}\,$yr \citep{MS03}, but 
recent observations of the proximity effect in high-redshift 
QSOs \citep{HC02,YL05} and the transverse
proximity effect in He II \citep{Jakobsen03} 
also imply a lifetime of $\tQ\sim10^{7}\,$yr. 
If the net and episodic lifetimes are similar and $\sim10^{7}\,$yr, quasars could
be produced by the relatively rare mergers of equal mass, gas rich galaxies. 
This mechanism has some good observational support \citep{Heckman84,HN92}, 
and is also favored theoretically, as major mergers can efficiently dissipate angular 
momentum and drive gas to the center of galaxies 
\citep[e.g.,][]{Hernquist89,BH91,BH96,MH96}. 
%This mechanism is currently favored, as it explains the accretion rates necessary
%to fuel quasars through the efficient stripping of angular momentum and driving
%of gas to the center of the galaxies (e.g. Hernquist 1989; Barnes \& Hernquist
%1991, 1996; Mihos \& Hernquist 1996)
%and the rapid decrease in their space density at low redshift 
%(Carlberg 1990; for a review see, e.g. Barnes \& Hernquist 1992).

Theoretical models of supermassive black hole evolution 
and its correlation with galaxy structure predict that, beyond a certain threshold,
the feedback energy of the black hole accretion expels nearby gas and shuts down 
the accretion phase \citep{CO97,CO01,SR98,WL03}.
However, these models do not predict the characteristic lifetime of the accretion phase 
prior to its self-termination, but usually take it to be an 
independent input parameter of the model. 
The input lifetime is either taken from observational estimates or assumed to be similar to
a characteristic timescale such as the dynamical time of the host galaxy disk or 
the $e$-folding time for Eddington-limited black hole growth
$t_{S}=M_{\rm BH}/\Mdot=4\times10^{8}\,\epsilon l\,$yr for accretion with
radiative efficiency $\epsilon=L/\Mdot c^{2}\sim0.1$ and $l=L/L_{Edd}\lesssim1$
\citep{Salpeter64}.

However, recent hydrodynamical simulations \citep{DSH05, SDH05b}, which
include black holes, associated gas accretion and feedback processes
during galaxy mergers, have shown that black hole growth is determined
by the gas supply and terminates abruptly when significant gas is
expelled as a result of the coupling of radiative feedback energy from
black hole accretion to the surrounding gas. This determines a
timescale for a strong accretion phase - i.e. a quasar phase for the
black hole - of $\sim 10^{8}\,{\rm yr}$ (for a Milky Way mass
system). This timescale is approximately equal to the gas inflow timescale, 
set by the time when strong torques are present in the central region, 
and is substantially shorter than the merger timescale of 
$\sim2\times10^{9}\,{\rm yr}$ \citep{Hernquist89,BH91,BH96}.  The
timescale for associated starbursts can be similarly understood as
dependent on large densities and thus related to the timescale of gas
inflow \citep{MH94,MH96}. In order to compare the quasar lifetime
predicted in these simulations with the observational
estimates of the quasar lifetime, mostly inferred from visible wavelength
observations, is necessary to take into account additional
physical mechanisms.

Previous theoretical models of supermassive black hole and quasar evolution 
assume that the observed quasar lifetime and the lifetime of the 
accretion phase are the same. However, the accretion rate is 
tied to the surrounding density of gas, and quasar activity occurs when 
material is efficiently channeled to the center of the galaxy onto the supermassive
black hole. Therefore, the time(s) of 
greatest accretion activity are likely to also have the largest obscuring column density, 
especially if the line-of-sight column to the black hole is dominated by the density in 
the central, most dense regions of the galaxy. The corresponding attenuation 
may render the quasar unobservable in some bands or lower its observed luminosity below the
quasar threshold, giving a shorter observed lifetime than intrinsic lifetime. 
Modeling this effect is necessary to estimate the intrinsic quasar 
lifetimes from observations, as well as for using theoretically motivated accretion
models to predict the quasar luminosity function and space density of present-day
supermassive black holes. Further, this effect can account for the presence of 
an obscured population of quasars which are missed by optical, UV, or soft X-ray
surveys but significantly contribute to the cosmic X-ray background \citep[and references therein]{BH05}.
Similarly, re-processing of quasar radiation by dust in a large 
obscuring column can account for and model observations of luminous and ultra-luminous
infrared galaxies, which show evidence of both merger activity 
and obscured AGN in their cores \citep[e.g.,][]{SM96,Komossa03}. 

In this paper, we propose a model of quasar lifetimes in which quasars
are created and fed in major mergers, but for a large part of the
accretion lifetime they are heavily obscured by the large gas
densities powering accretion. Eventually, feedback from the accretion
energy drives away gas, creating a brief window in which the central
object is observable as a quasar, until accretion levels drop below
quasar thresholds. To test this, we determine quasar lifetimes \tq\
from a simulation of a major merger of gas-rich galaxies, calculating
the effects of obscuration. Using a model for black hole accretion and
associated feedback developed by \citet{SDH05b, DSH05}, the
simulation allows us to determine unambiguously the intrinsic lifetime
of a strong accretion phase in the merger. By simultaneously tracing
the column densities along different lines-of-sight from the
simulation supermassive black hole, here we calculate the observed
luminosity in visible wavelengths, and determine a lifetime over
which the AGN would be observed as a quasar, with luminosity above
some threshold. We find that there is a significant difference between
the intrinsic lifetime and the observable value, and find that
our predicted observed lifetimes agree well with the observational estimates, with 
a substantially longer intrinsic lifetime.

\section{The Simulations\label{sec:sim}}
The simulations used the GADGET-2 code, a new version of the parallel
TreeSPH code GADGET \citep{SYW01}. It uses an entropy-conserving
formulation of SPH \citep{SH02}, and includes radiative cooling,
heating by a UV background, and a sub-resolution model of a two-phase
structure of the dense ISM to describe star formation and supernova
feedback \citep{SH03}. This sub-resolution model gives an effective
equation of state which includes pressure feedback from supernovae
heating, and allows us to stably evolve even massive pure gaseous
disks (see, e.g. Robertson et al. 2004). The methodology of accretion,
feedback, and galaxy generation is described in detail in
\citet{SDH05b}.  In particular, in this model, supermassive black
holes (BHs) are represented by ``sink'' particles that accrete gas
from their local environment, with an accretion rate $\Mdot$ estimated
from the local gas density and sound speed using a
Bondi-Hoyle-Lyttleton parameterization with an imposed upper limit
equal to the Eddington rate. The bolometric luminosity of the BH
particle is then $\Lbol=\dEdt$, with $\epsilon=0.1$ the
accretion efficiency. We further assume that a small fraction ($5\%$)
of $\Lbol$ couples dynamically to the surrounding gas, and this
feedback is injected as thermal energy. This
fraction is a free parameter, determined in \citet{DSH05} by fitting
to the $M_{\rm BH}$-$\sigma$ relation.  We do not attempt to resolve
the small-scale accretion dynamics near the black hole, but instead
assume that the time-averaged accretion can be estimated from the gas
properties on the scale of our spatial resolution ($\lesssim100$\,pc).

We generate two stable, isolated disk galaxies, each with 
an extended dark matter halo with a \citet{Hernquist90} profile,
an exponential gas disk, and a bulge. Our simulation is one of the
series described in detail in \citet{SDH05a}, 
with virial velocity $V_{\rm vir}=160\,{\rm km\,s^{-1}}$, 
a fiducial choice with a rotation curve and mass similar to the Milky Way.
We begin our simulation with pure gaseous disks, which may better correspond to 
the high-redshift galaxies in which most quasars are observed. 
%The galaxies have mass $M_{\rm vir}=V_{\rm vir}^{3}/(10GH_{0})$, with the 
%baryonic disk having mass fraction $m_{\rm d}=0.041$, the bulge $m_{\rm b}=0.0136$, 
%and the rest of the mass in dark matter with a concentration parameter $9.0$. 
%The disk-scale length is computed based on an assumed 
%spin parameter $\lambda=0.033$, and the scale-length of the bulge is set to $0.2$ times this.
Each galaxy is initially 
composed of 168000 dark matter halo particles, 8000 bulge particles, 
24000 gaseous disk particles, and one BH particle, with a small initial seed mass of 
$10^{5}M_{\sun}$. Given these choices, the dark matter, gas, and star particles are all of
roughly equal mass, and central cusps in the dark matter and bulge profiles 
\citep{Hernquist90} are reasonably well resolved. The galaxies are then set to collide in
a pure prograde encounter with zero orbital energy and a pericenter separation of $7.1\,{\rm kpc}$.

\section{Column Densities \&\ Quasar Attenuation\label{sec:NH}}
We calculate the column density between a black hole and a hypothetical observer 
from the simulation outputs spaced every 10 Myr before and after the merger and every 
5 Myr during the merger (1.14 - 1.71 Gyr). We first generate 
$\sim1000$ radial lines-of-sight (rays), 
each with its origin at the black hole particle location and with directions uniformly spaced in 
solid angle $d\cos{\theta}\,d\phi$. For each ray, we then begin at the origin, calculate 
and record the local gas properties using GADGET, and then move a distance along the ray 
$\Delta r=\eta h_{\rm sml}$, where $\eta \leq 1$ and $h_{\rm sml}$ is the local 
SPH smoothing length. The process is repeated until a ray is sufficiently far from its origin
($\gtrsim 100$ kpc). The gas properties along a given ray can then be integrated to give the 
line-of-sight column density and mean metallicity. 
We test different values of $\eta$ and find that gas properties along a ray
converge rapidly and change smoothly for $\eta=0.5$ and smaller. We similarly test different
numbers of rays and find that the distribution of line-of-sight properties
converges for $\gtrsim 100$ rays. 

Given the local gas properties, we use the GADGET 
two-phase equilibrium model of the ISM described in \citet{SH03} to calculate the local
mass fraction in ``hot'' (diffuse) and ``cold'' (molecular and HI cloud) phases of 
dense gas, and assuming
pressure equilibrium between the two phases we obtain the density of the local hot and cold 
phase gas and the corresponding volume filling factors. These values correspond roughly to the 
fiducial values of \citet{MO77}. Using only the hot-phase density allows us to place 
an effective lower limit on the column density along a particular line of sight, as it assumes
a ray passes only through the diffuse ISM, with $\gtrsim 90\%$ of the mass of the dense ISM 
concentrated in cold-phase ``clumps.'' Given the small volume filling factor 
($<0.01$) and cross section of such clouds, we expect that the majority of sightlines will 
pass only through the ``hot-phase'' medium, with rare outliers dominated by single
clouds along the line of sight (covering fraction $\lesssim1\%$).

Using $\Lbol=\dEdt$, we model the form of the intrinsic quasar continuum SED 
following \citet{Marconi04}, based on optical through hard X-ray observations
\citep[e.g.,][]{Elvis94,George98,VB01,Perola02,Telfer02,Ueda03,VBS03}.
%with a broken power law in the optical-UV \citep{Telfer02,VB01}, 
%and a power law \citep{George98,Perola02} plus reflection component \citep{Ueda03} in the 
%X-ray, scaled with a luminosity-dependent $\alpha_{OX}$ \citep{VBS03}. 
This gives a B-band luminosity 
$\log{(\LB)}=0.80-0.067\mathcal{L}+0.017\mathcal{L}^{2}-0.0023\mathcal{L}^{3}$, 
where $\mathcal{L} = \log{(\Lbol/L_{\sun})} - 12$, and we take 
$\lambda_{B}=4400\,$\AA . 
%For the optical and UV, we then use a gas-to-dust ratio to
%determine the extinction along a given line of sight at a particular frequency. 
We then use a gas-to-dust ratio to determine the extinction along a given line of sight 
at this frequency.
Observations suggest that the majority of the population of reddened quasars have reddening curves
similar to that of the Small Magellenic Cloud (SMC) \citep{Hopkins04}, which 
has a gas-to-dust ratio lower than the Milky Way by approximately the same factor as its 
metallicity \citep{Bouchet85}. We consider both a gas-to-dust ratio equal to that of the Milky Way, 
$(A_{B}/N_{H})_{\rm MW}=8.47\times10^{-22}\,{\rm cm^{2}}$, and a gas-to-dust ratio scaled by metallicity, 
$A_{B}/N_{H} = (Z/0.02)(A_{B}/N_{H})_{\rm MW}$. For both cases we use 
%$(A_{V}/N_{H})_{\rm MW}=6.39\times10^{-22}\,{\rm cm^{2}}$, and a gas-to-dust ratio scaled by metallicity, 
%$A_{V}/N_{H} = (Z/0.02)(A_{V}/N_{H})_{\rm MW}$. For both cases we use 
the SMC-like reddening curve of \citet{Pei92}. 
%For comparison, we 
%calculate extinction in X-ray frequencies (0.03-10 keV) using the 
%photoelectric absorption cross sections of \citet{MM83}, similarly scaled by metallicity.
We do not perform a full radiative transfer calculation but defer this to a future paper, 
and therefore do not model scattering or re-processing of radiation by dust.

%\begin{center}
\begin{figure}
    \plotone{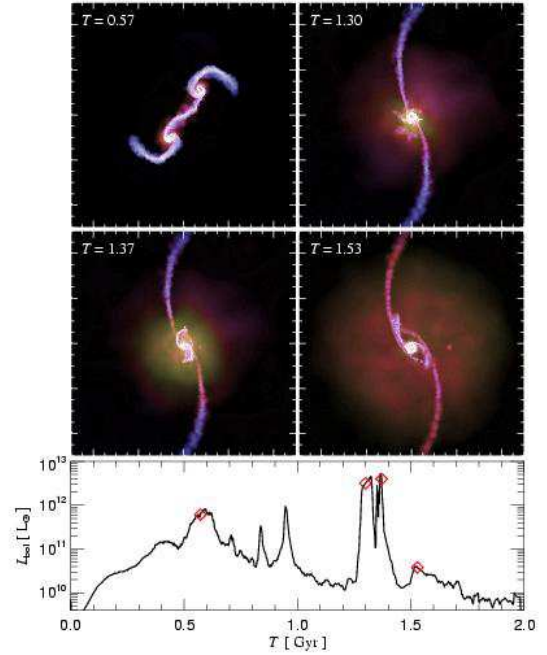}
    \caption{The projected gas density is shown in 
    boxes $100\,h^{-1}\,{\rm kpc}$ on a side, color-coded by temperature (blue to red).
    The bolometric luminosity of the central black hole(s)
    is shown in the lower panel, with diamonds marking the times shown above. 
    Bolometric luminosities prior to the 
    merger are the sum of the two individual black hole luminosities.
    \label{fig:showGADGET}}
\end{figure}
\begin{figure}
    %\plotone{f2.eps}
    \centering
    \includegraphics[width=3.7in]{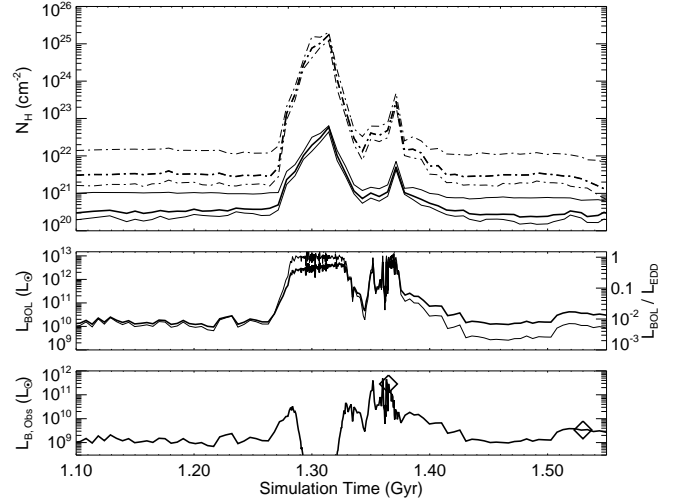}
    \caption{Upper panel: Thick contours plot the median column density \NH\ as 
    a function of simulation time, with thin contours at 25\% and 75\% inclusion levels. 
    Solid contours represent the density of the 
    ``hot-phase'' ISM, %calculated as described in \S\,3, %\ref{sec:NH}
    dashed contours the total simulation density. 
    Middle panel: Bolometric luminosity of the simulation black hole, $\Lbol=\dEdt$ (thick), 
    and ratio of bolometric to Eddington luminosity, $l\equiv\Lbol/L_{Edd}$ (thin). Values
    are shown for each simulation timestep.
    Lower panel: Observed B-band luminosity calculated given the median ``hot-phase'' ISM 
    density. Diamonds mark times shown in Figure\,\ref{fig:showGADGET}.
    \label{fig:NH.vs.time}}
\end{figure}
\begin{figure}
    %\plotone{f3.eps}
    \centering
    \includegraphics[width=3.7in]{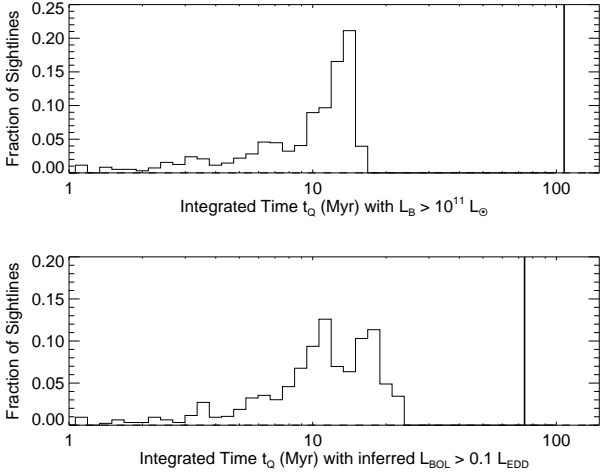}
    \caption{Upper panel: Histograms of the observed quasar lifetime \tq, defined as an observed
    B-band luminosity $\LBo\ge\Lcut{11}$, for the simulation along different
    lines-of-sight to the black hole, given the ``hot-phase''
    ISM density. The thick line shows the lifetime if 
    attenuation is ignored, $\tQ\approx 113$\,Myr.
    Lower panel: Same as above, but with lifetime \tq\ 
    defined as an inferred bolometric luminosity
    (estimated from $\LBo$) $\Lbol>0.1\,L_{Edd}$.
    \label{fig:lifetime.hists}}
\end{figure}
\begin{figure}
    %\plotone{f4.eps}
    \centering
    \includegraphics[width=3.7in]{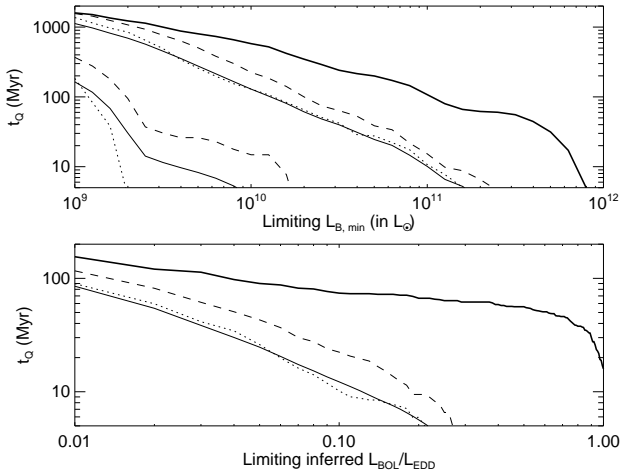}
    \caption{Upper panel: Quasar lifetimes as a function of $\LBm$ from the simulation.
    The thick line shows the lifetime if attenuation is ignored (time with observed
    B-band luminosity $\LBo\geq \LBm$).
    The lower three lines use the total ISM density, middle three the ``hot-phase'' density. 
    Solid lines are the integrated lifetime weighted by the fraction of sightlines along which 
    the definition is satisfied. Dashed lines are the lifetime for which $\ge10\%$ of sightlines
    meet the definition, dotted lines for which $\ge50\%$ meet the definition. 
    Lower panel: Same as above, but with \tq\ defined as an inferred bolometric luminosity
    (estimated from $\LBo$) $\Lbol>l\,L_{Edd}$. The total ISM density attenuates the quasar
   below the limits shown, so only the ``hot-phase'' results are shown. 
   \label{fig:lifetime.vs.L}}
\end{figure}
%\end{center}

\section{Results\label{sec:results}}
Figure \ref{fig:showGADGET} shows the simulation SPH particles at four representative
times during the run, with the bolometric luminosity of the supermassive black hole 
as a function of time below. 
After the first passage (upper left), there is an extended period of strong accretion, but central gas
densities are very large and the intrinsic quasar luminosity will be attenuated well below 
our quasar threshold of $\LBo\geq10^{11}\,L_{\sun}$ ($M_{\rm B}\lesssim-23$). 
Once the merger begins (upper right), 
the intrinsic luminosity peaks as gas is channeled to the merging cores. However, this 
also results in very large columns which similarly obscure the quasar. After a short time
(lower left), feedback from the quasar clears out the gas in the central regions and 
the object may be observable as a quasar. Shortly after the merger (lower
right), gas has been consumed by star formation and accretion or expelled from the center
and densities have dropped well below the levels needed to fuel quasar activity. 

Figure \ref{fig:NH.vs.time} shows the bolometric luminosity of and column 
densities to the supermassive
black hole as a function of simulation time, for the period shortly before and
after the merger ($\approx1.27\,{\rm Gyr}$). 
The median column densities and 25\%-75\% contours are shown
for the total average simulation density, and the calculated
hot-phase density, as described in \S\,3.%\ref{sec:NH}. 
We note that the metallicity-weighted
column densities show little difference from the non-weighted columns. 
This is because the column is dominated by large densities in the 
central, most dense regions, where star formation rates (and thus metallicities) are large. 
The bolometric luminosity
shows a strong rise above $10^{12}\,L_{\sun}$ during the merger, but the column 
densities simultaneously become very large, up to $\sim10^{25}\,{\rm cm^{-2}}$
for the total column density and 
$\sim10^{23}\,{\rm cm^{-2}}$ for the hot-phase column. These values correspond to the
large columns seen towards obscured quasars and suggest that the quasar will not
be observable at visible wavelengths for a significant fraction of this period. 

We quantify the resulting quasar lifetimes in Figure \ref{fig:lifetime.hists}. For each
sightline, an observed lifetime is determined as the integrated time in the simulation 
during which the given sightline sees a B-band luminosity above a given threshold, 
$\LBo\geq10^{11}\,L_{\sun}$. The histogram plots the fraction of sightlines 
with a particular observed quasar lifetime \tq, spaced in logarithmic bins. Lifetimes are
calculated using the metallicity-weighted hot-phase density. Using the total (cold-phase)
column density attenuates the quasar below this limit along all sightlines at all times. 
The line at $\tQ\approx1.1\times10^{8}\,$yr shows the intrinsic lifetime along all sightlines if 
attenuation is ignored. We note that below 5 Myr (the simulation output frequency)
our estimates of \tq\ become 
uncertain owing to the effects of variability and our inability to resolve the local small-scale
physics of the ISM. However, the more reliable estimates of a typical observation, from the 
hot-phase column density, lie well above this limit. The majority of such sightlines 
see lifetimes in the range $10-20\,$Myr, in good agreement with observations suggesting
lifetimes $\sim10^{7}\,$yr. The longest observable lifetimes in this case are approximately
$20\,$Myr, $\sim1/6$ of the intrinsic strong accretion-phase lifetime. 
We also show the observed lifetime with an inferred %\ref{sec:NH}) 
bolometric luminosity (calculated from the observed
$\LBo$ using the bolometric corrections defined in \S\,3)
greater than $0.1\,L_{Edd}$, where $L_{Edd}$ is the Eddington luminosity for the 
black hole mass at a given time ($L_{Edd}=6\times10^{12}\,L_{\sun}$ for the final 
black hole mass $1.9\times10^{8}\,M_{\sun}$), 
and find similar lifetimes. 

In Figure \ref{fig:lifetime.vs.L} we plot the quasar lifetime \tq\ as a function
of the limiting B-band luminosity $\LBm$. The uppermost curve shows the intrinsic lifetime
$\tQ^{\,i}$, the total time the intrinsic $\LB\geq \LBm$.
The bottom curves show the integrated time that the observed B-band luminosity
meets this criterion, using the total metallicity-weighted column density of the simulation. The 
solid curve shows the integrated time, weighted by the fraction of sightlines $f$ along which the
above definition is met [$\tQ=\int{f(\LBo\geq \LBm)\,dt}$]. The dotted curve shows
the integrated time during which $\geq 50\%$ of sightlines meet the criterion, and the 
dashed curve the integrated time during which $\geq 10\%$ of sightlines meet the criterion.
The middle curves show the same, using the hot-phase density. In all cases, changing 
the definition of \tq\ based on the fraction of sightlines with a given column does not 
significantly change \tq. However, \tq\ is, unsurprisingly, a strong function of $\LBm$.
The observed \tq\ is significantly smaller than the intrinsic lifetime 
for all $\LBm>10^{9}\,L_{\sun}$, and the ratio $\tQ/\tQ^{\,i}$ decreases with 
increasing $\LBm$. We also plot the lifetime as a function of $l$, the ratio of the
inferred bolometric luminosity to the Eddington luminosity, and find a similar trend. 

\section{Conclusions\label{sec:conclusions}}
We find that incorporating the effects of obscuration in a galaxy merger simulation 
gives observed quasar lifetimes of $\sim10^{7}\,{\rm yr}$, in good agreement with
observational estimates. 
For a significant fraction of the intrinsic lifetime for quasar activity, the object
is heavily obscured, and attenuated well below observable limits in the B-band. 
%However, for much of this regime the X-ray attenuation is much weaker, 
%and the object may be observable at these frequencies. 
The significant difference between the observed and intrinsic lifetimes 
is critical for modeling quantities dependent on observed lifetimes, such as the 
quasar luminosity function.  This distinction must also be accounted for in comparing these quantities 
with others which depend on the much larger intrinsic lifetimes, such as the
present-day population of supermassive black holes or the cosmic X-ray background spectrum. 
Furthermore, our model naturally predicts that the observed
lifetime is itself a function of the observed frequency, an effect that must be considered in 
extending an observed quasar luminosity function to other wavelengths 
or using observational constraints of quasar lifetimes
to study the infrared and X-ray backgrounds. Estimates of the fraction of nuclear or starbursting
galaxies with nuclear activity must similarly combine multiwavelength observations as optical or 
IR observations alone may miss a large fraction of AGN. 

Our model predicts that self-regulating feedback processes in galaxy mergers 
reproduce observed quasar lifetimes naturally, in the final stages of strong accretion 
as feedback removes nearby gas. In determining the lifetime and evolution of 
accretion these feedback processes must be taken into account. 
Major merger scenarios using this model reproduce observed lifetimes of
quasars with approximately equivalent episodic and total observed lifetimes. 
The population of obscured quasars is a natural consequence of the model, 
not as an independent population but as a stage in the ``standard'' evolution 
of quasars over their lifetime, before feedback can clear sufficient material 
to render the quasar visible. 

As the black hole growth and column densities are strong functions of the 
galaxy sizes and merger conditions, we defer a full treatment of the statistics of 
quasar lifetime distributions as a function of these parameters to a following paper. 
However, the relationship between observed and intrinsic quasar lifetimes should 
remain qualitatively similar under a wide range of initial conditions, as it is determined by 
feedback mechanisms in the final stages of black hole growth. Using this model, a wider 
examination of simulations will allow for a consistent prediction of quasar lifetimes as a 
function of observed wavelength, redshift, and host galaxy properties. 

\acknowledgments
This work was supported in part by NSF grants ACI
96-19019, AST 00-71019, AST 02-06299, and AST 03-07690, and NASA ATP
grants NAG5-12140, NAG5-13292, and NAG5-13381.
The simulations
were performed at the Center for Parallel Astrophysical Computing at the 
Harvard-Smithsonian Center for Astrophysics.


\begin{thebibliography}{}
\bibitem[Bajtlik, Duncan, \&\ Ostriker(1988)]{BDO88}
Bajtlik, S., Duncan, R. C., \&\ Ostriker, J. P. 1988, \apj, 327, 570
\bibitem[Barnes \&\ Hernquist(1991)]{BH91}
Barnes, J. E. \&\ Hernquist, L. 1991, \apj, 370, L65
%\bibitem[Barnes \&\ Hernquist(1992)]{BH92}
%Barnes, J. E. \&\ Hernquist, L. 1992, \araa, 30, 705
\bibitem[Barnes \&\ Hernquist(1996)]{BH96}
Barnes, J. E. \&\ Hernquist, L. 1996, \apj, 471, 115
%\bibitem[Bennert \etal(2002)]{Bennert02}
%Bennert, N., Falcke, H., Schulz, H., Wilson, A. S., \&\ Wills, B. J. 2002, \apj, 574, L105
\bibitem[Bouchet et al.(1985)]{Bouchet85} 
Bouchet, P., Lequeux, J., Maurice, E., Prevot, L., \&\ 
Prevot-Burnichon, M.~L.\ 1985, \aap, 149, 330 
%\bibitem[Carlberg(1990)]{Carlberg90}
%Carlberg, R. G. 1990, \apj, 350, 505
\bibitem[Brandt \&\ Hasinger(2005)]{BH05}
Brandt, W.N., Hasinger G., 2005, A. Rev.A.\& A, in press
\bibitem[Ciotti \&\ Ostriker(1997)]{CO97}
Ciotti, L. \&\ Ostriker, J. P. 1997, \apj, 487, L105
\bibitem[Ciotti \&\ Ostriker(2001)]{CO01}
Ciotti, L. \&\ Ostriker, J. P. 2001, \apj, 551, 131
%\bibitem[Ciotti \&\ van Albada(2001)]{CV02}
%Ciotti, L. \&\, van Albada T.S., 2001, \apjl, 552, L13
\bibitem[Di Matteo \etal(2003)]{DCSH03} 
Di Matteo, T., Croft, ~R.~A.~C, Springel, V., \&\ Hernquist, L. 2003, \apj, 593, 56
\bibitem[Di Matteo, Springel, \&\ Hernquist(2005)]{DSH05}
Di Matteo, T., Springel, V., \&\ Hernquist, L. 2005, Nature, in press
\bibitem[Elvis et al.(1994)]{Elvis94} 
Elvis, M., et al.\ 1994, \apjs, 95, 1 
\bibitem[Ferrarese \&\ Merritt(2000)]{FM00}
Ferrarese, L. \&\ Merritt, D. 2000, \apjl, 539, L9
\bibitem[Gebhardt \etal(2000)]{Gebhardt00}
Gebhardt, K., Bender, R., Bower, G. \etal\ 2000, \apjl, 539, L13
\bibitem[George \etal(1998)]{George98}
George, I. M., Turner, T. J., Netzer, H., Nandra, K., Mushotzsky, R. F., \&\ 
Yaqoob, T. 1998, \apjs, 114, 73
\bibitem[Granato \etal(2004)]{Gran04} Granato G.L., De Zotti G., Silva L., Bressan A., Danese L., 
2004, \apj, 600, 580
\bibitem[Grazian \etal(2004)]{Grazian04}
Grazian, A., Negrello, M., Moscardini, L., Cristiani, S., Haehnelt, M. G., 
Matarrese, S., Omizzolo, A., \&\ Vanzella, E. 2004, \aj, 127, 592
\bibitem[Haehnelt, Natarajan, \&\ Rees(1998)]{HNR98}
Haehnelt, M. G., Natarajan, P., \&\ Rees, M. J. 1998, \mnras, 300, 817
\bibitem[Haiman \&\ Cen(2002)]{HC02}
Haiman, Z. \&\ Cen, R. 2002, \apj, 578, 702
\bibitem[Heckman et al.(1984)]{Heckman84} 
Heckman, T.~M., Bothun, G.~D., Balick, B., \&\ Smith, E.~P.\ 1984, \aj, 89, 958 
\bibitem[Hernquist(1989)]{Hernquist89}
Hernquist, L. 1989, Nature, 340, 687
\bibitem[Hernquist(1990)]{Hernquist90}
Hernquist, L. 1990, \apj, 356, 359
%\bibitem[Hopkins \etal(2004)]{Hopkins04}
%Hopkins, P. F., Strauss, M. A., Hall, P. B., Richards, G. T., Cooper, A. S., 
%Schneider, D. P., Vanden Berk, D. E., Jester, S., Brinkmann, J., \&\ Szokoly, G. P.
%2004, \aj, 128, 1112
\bibitem[Hopkins et al.(2004)]{Hopkins04} 
Hopkins, P.~F., et al.\ 2004, \aj, 128, 1112 
\bibitem[Hutchings \&\ Neff(1992)]{HN92} 
Hutchings, J.~B., \&\ Neff, S.~G.\ 1992, \aj, 104, 1 
\bibitem[Jakobsen \etal(2003)]{Jakobsen03}
Jakobsen, P., Jansen, R. A., Wagner, S., \&\ Reimers, D. 2003, \aap, 397, 891
\bibitem[Kauffmann \&\ Haehnelt(2000)]{KH00}
Kauffmann, G. \&\ Haehnelt, M. 2000, \mnras, 311, 576
\bibitem[Komossa \etal(2003)]{Komossa03}
Komossa, S., Burwitz, V., Hasinger, G., Predehl, P., Kaastra, J. S., \&\ Ikebe, Y.
2003, \apj, 582, L15
%\bibitem[Lequeux et al.(1979)]{Lequeux79} 
%Lequeux, J., Peimbert, 
%M., Rayo, J.~F., Serrano, A., \&\ Torres-Peimbert, S.\ 1979, \aap, 80, 155 
\bibitem[Marconi \etal(2004)]{Marconi04}
Marconi, A., Risaliti, G., Gilli, R., Hunt, L. K., Maiolino, R., \&\ Salvati, M. 
2004, \mnras, 351, 169
\bibitem[Martini \&\ Schneider(2003)]{MS03}
Martini, P. \&\ Schneider, D.~P.\ 2003, \apjl, 597, L109 
\bibitem[Martini(2004)]{Martini04}
Martini, P. 2004, in Carnegie Obs. Astrophys. Ser. 1, Coevolution of Black Holes
and Galaxies, ed. L.C. Ho (Cambridge: Cambridge Univ. Press), 170
\bibitem[McKee \&\ Ostriker(1977)]{MO77}
McKee, C. F. \&\ Ostriker, J. P. 1977, \apj, 218, 148
\bibitem[Mihos \&\ Hernquist(1994)]{MH94} 
Mihos, J.~C., \&\ Hernquist, L.\ 1994, \apj, 437, 611 
\bibitem[Mihos \&\ Hernquist(1996)]{MH96}
Mihos, J.C. \&\ Hernquist, L. 1996, \apj, 464, 641
%\bibitem[Morrison \&\ McCammon(1983)]{MM83}
%Morrison, R. \&\ McCammon, D. 1983, \apj, 270, 119
\bibitem[Pei(1992)]{Pei92} 
Pei, Y. C. 1992, \apj, 395, 130
\bibitem[Perola \etal(2002)]{Perola02} 
Perola, G.~C., Matt, G., Cappi, M., Fiore, F., Guainazzi, M., 
Maraschi, L., Petrucci, P.~O., \&\ Piro, L.\ 2002, \aap, 389, 802
\bibitem[Porciani, Magliocchetti, \&\ Norberg(2004)]{PMN04}
Porciani, C., Magliocchetti, M., \&\ Norberg, P. 2004, \mnras, 355, 1010
\bibitem[Robertson \etal(2004)]{Robertson04}
Robertson, B., Yoshida, N., Springel, V., \&\ Hernquist, L. 2004, \apj, 606, 32
\bibitem[Salpeter(1964)]{Salpeter64}
Salpeter, E. E. 1964, \apj, 140, 796
\bibitem[Sanders \&\ Mirabel(1996)]{SM96}
Sanders, D. B. \&\ Mirabel, I. F. 1996, \araa, 34, 749
\bibitem[Silk \&\ Rees(1998)]{SR98}
Silk, J. \&\ Rees, M. J. 1998, \aap, 331, L1
%\bibitem[Sokasian \etal(2002)]{Sokasian02}
%Sokasian, A., Abel, T., \& Hernquist, L. 2002, \mnras, 332, 601
\bibitem[Soltan(1982)]{Soltan82}
Soltan, A. 1982, \mnras, 200, 115
\bibitem[Springel, Di Matteo, \&\ Hernquist(2005a)]{SDH05a}
Springel, V., Di Matteo, T., \&\ Hernquist, L. 2005a, \apjl, submitted, (astro-ph/0409436)
\bibitem[Springel, Di Matteo, \&\ Hernquist(2005b)]{SDH05b}
Springel, V., Di Matteo, T., \&\ Hernquist, L. 2005b, \mnras, submitted, (astro-ph/0411108)
\bibitem[Springel \&\ Hernquist(2002)]{SH02} 
Springel, V. \&\ Hernquist, L. 2002, \mnras, 333, 649
\bibitem[Springel \&\ Hernquist(2003)]{SH03} 
Springel, V. \&\ Hernquist, L. 2003, \mnras, 339, 289
\bibitem[Springel, Yoshida, \&\ White(2001)]{SYW01}
Springel, V., Yoshida, N., \&\ White, S. D. M. 2001, New Astronomy, 6, 79
\bibitem[Telfer \etal(2002)]{Telfer02}
Telfer, R.~C., Zheng, W., Kriss, G.~A., \&\ Davidsen, A.~F.\ 2002, \apj, 565, 773
%\bibitem[Tremaine \etal(2002)]{Tremaine02}
%Tremaine, S., et al. 2002, \apj, 574, 740
\bibitem[Ueda \etal(2003)]{Ueda03} 
Ueda, Y., Akiyama, M., Ohta, K., \&\ Miyaji, T.\ 2003, \apj, 598, 886
\bibitem[Vanden Berk \etal(2001)]{VB01} 
Vanden Berk, D.~E., \etal 2001, \aj, 122, 549 
\bibitem[Vignali \etal(2003)]{VBS03} 
Vignali, C., Brandt, W.~N., \&\ Schneider, D.~P.\ 2003, \aj, 125, 433
\bibitem[Wyithe \&\ Loeb(2002)]{WL02}
Wyithe, J. S. B. \&\ Loeb, A. 2002, \apj, 581, 886
\bibitem[Wyithe \&\ Loeb(2003)]{WL03}
Wyithe, J. S. B. \&\ Loeb, A. 2003, \apj, 595, 614
\bibitem[Yu \&\ Lu(2004)]{YL04}
Yu, Q. \&\ Lu, Y. 2004, \apj, 602, 603
\bibitem[Yu \&\ Lu(2005)]{YL05}
Yu, Q. \&\ Lu, Y. 2005, \apj, accepted, in press (astro-ph/0411098)
\bibitem[Yu \&\ Tremaine(2002)]{YT02}
Yu, Q. \&\ Tremaine, S. 2002, \mnras, 335, 965
\end{thebibliography}
\end{document}